\documentclass[final]{aipproc}
\layoutstyle{6x9}
\usepackage{pstricks,epsfig}

\psset{unit=5mm,dash=2mm 1mm,dotsep=0.5mm,hatchsep=2pt,labelsep=3pt}

\newcommand{\be}{\begin{eqnarray}}
\newcommand{\ee}{\end{eqnarray}}
\newcommand{\ba}{\begin{array}}
\newcommand{\ea}{\end{array}}
\newcommand{\bt}{\begin{tabular}}
\newcommand{\et}{\end{tabular}}

\begin{document}

\title{Relativistic Bound States}

\author{Axel Weber\footnote{
e--mail: axel@itzel.ifm.umich.mx}}{
address={Instituto de F\'{\i}sica y Matem\'aticas, Universidad Michoacana
de San Nicol\'as de Hidalgo, \\
Edificio C--3, Ciudad Universitaria, A. P. 2--82, 58040 Morelia,
Michoac\'an, Mexico}
}

\begin{abstract}
In this contribution, I will give a brief survey of present techniques 
to treat the bound state problem in relativistic quantum field
theories. In particular, I will discuss the Bethe--Salpeter equation,
various quasi--potential equations, the Feynman--Schwinger representation,
and similarity transformation methods for Hamiltonian approaches
in light--front quantization. Finally, I will comment on a related
similarity transformation in the usual equal--time quantized theory.
\end{abstract}

\maketitle

One of the most striking and ubiquitous phenomena in nature is the
binding of matter. It is found on essentially all length scales, from 
galaxies to quarks, and is even speculated to be crucial for the
description of physics much below the length scale of hadronic physics.
We currently have a good understanding of the binding mechanism in
crystals, molecules, and atoms. The non--relativistic
Schr\"odinger equation provides a satisfactory description at these scales,
and small corrections due to relativistic and virtual particle creation
effects can accurately be treated in perturbation theory. However, at
smaller length scales relativistic effects become increasingly important,
and the description through the non--relativistic Schr\"odinger equation is 
not accurate enough. It is hence inevitable to employ quantum field
theory (QFT), the best currently known description of microscopic physics.

In the following, I will give a brief survey of the most popular 
equations which have been employed for bound state calculations in
relativistic QFT. Attention will be restricted to equations which are rooted
in field theory, in particular, I will leave out the generalizations 
of relativistic one--particle quantum mechanics to two or more particles,
as well as equations with phenomenological input. Equations that are derived 
from field theory can naturally be divided into two
classes, those based on the manifestly covariant Lagrangian, 
path--intregal formulation of QFT, and those built on a Hamiltonian
formulation in a perturbatively defined Fock space. 

\begin{figure}
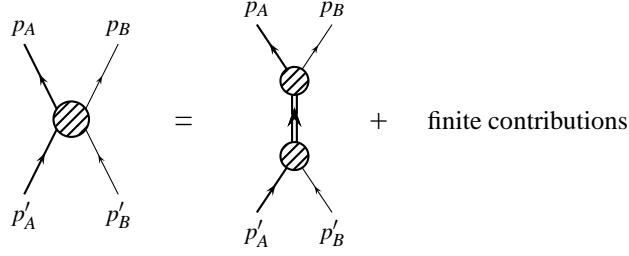

\pspicture[0.48](-1,-0.3)(4,6)
\psline(0.5,4)(0,5)
\psline{->}(1,3)(0.4,4.2)
\uput[90](0,5){$\scriptstyle p_A$}
\psline(0.5,2)(1,3)
\psline{->}(0,1)(0.55,2.1)
\uput[-90](0,1){$\scriptstyle p_A'$}
\psline[linewidth=.4pt](2,4)(2.5,5)
\psline[linewidth=.4pt]{->}(1.5,3)(2.07,4.15)
\uput[90](2.5,5){$\scriptstyle p_B$}
\psline[linewidth=.4pt](2,2)(1.5,3)
\psline[linewidth=.4pt]{->}(2.5,1)(1.97,2.05)
\uput[-90](2.5,1){$\scriptstyle p_B'$}
\pscircle[fillstyle=hlines*,fillcolor=white](1.25,3){0.5}
\endpspicture
= 
\pspicture[0.48](-1.25,0.2)(3.25,7.5)
\psline(0.75,5.75)(0.25,6.5)
\psline{->}(1.25,5)(0.59,6)
\uput[90](0.25,6.5){$\scriptstyle p_A$}
\psline(0.65,2.1)(1.25,3)
\psline{->}(0.25,1.5)(0.75,2.25)
\uput[-90](0.25,1.5){$\scriptstyle p_A'$}
\psline[doubleline=true](1.25,4.1)(1.25,5)
\psline[doubleline=true,arrowscale=0.8]{->}(1.25,3)(1.25,4.3)
\psline[linewidth=.4pt](1.75,5.75)(2.25,6.5)
\psline[linewidth=.4pt]{->}(1.25,5)(1.88,5.95)
\uput[90](2.25,6.5){$\scriptstyle p_B$}
\psline[linewidth=.4pt](1.85,2.1)(1.25,3)
\psline[linewidth=.4pt]{->}(2.25,1.5)(1.78,2.2)
\uput[-90](2.25,1.5){$\scriptstyle p_B'$}
\pscircle[fillstyle=hlines*,fillcolor=white](1.25,3){0.4}
\pscircle[fillstyle=hlines*,fillcolor=white](1.25,5){0.4}
\endpspicture
+ \quad {\footnotesize finite contributions}
\caption{The contribution of an $AB$--bound state to the 4--point function
for the scattering of particles $A$ and $B$.}
\label{fig:bspole}
\end{figure}

In the Lagrangian approach to QFT, which is particularly convenient for
the calculation of transition amplitudes, a bound state of two constituents
appears as a pole in the 2--particle scattering amplitude, or of the
field--theoretic 4--point correlation function. The pole is interpreted
as the pole in the propagator of the bound state as illustrated in Fig.\
\ref{fig:bspole}, and its position determines the bound state mass (via the
usual pole mass definition).
From the figure it is clear that near the pole the 4--point function has
the form
\be
i \, \frac{\psi (p_A, p_B) \bar{\psi} (p_A', p_B')}
{P^2 - M^2 + i \epsilon}
\ee
where $P = p_A + p_B = p_A' + p_B'$ is the total momentum, and finite
contributions have been suppressed. The function
\be
\psi(p_A, p_B) = \langle 0 | T [\phi_A (p_A) \phi_B (p_B) ] | \psi \rangle
\ee
is the so--called Bethe--Salpeter wave function and encodes the structure
of the bound state $| \psi \rangle$.

It is a well--known fact that no contribution of finite order in perturbation
theory, or, diagrammatically speaking, no finite number of Feynman diagrams, 
can ever give a bound state pole in the 4--point function. 
Hence it is imperative in the
Lagrangian approach to sum up an infinite subset of diagrams. It is this
property that makes the calculation of bound states a non--perturbative
problem in QFT, and the choice of an appropriate subset of diagrams and its
summation present formidable problems in practice.

\begin{figure}
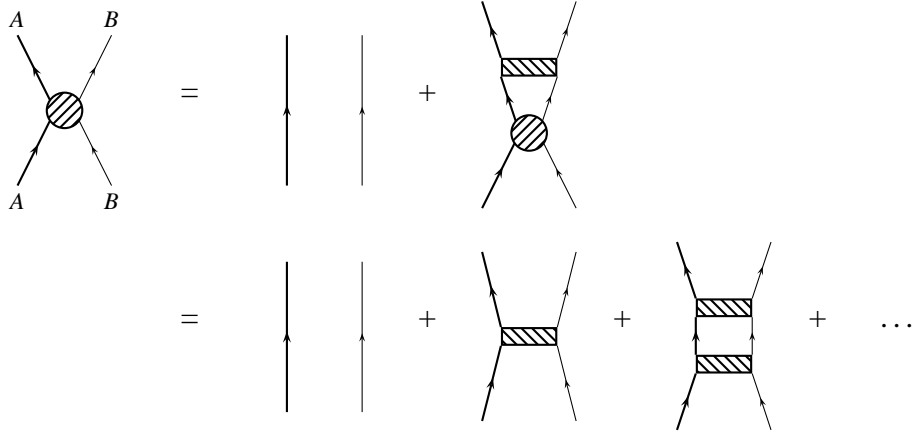

\bt{rcl}
\pspicture[0.48](-1,0.7)(3.5,6)
\psline(0.5,4)(0,5)
\psline{->}(1,3)(0.4,4.2)
\uput[90](0,5){$\scriptstyle A$}
\psline(0.5,2)(1,3)
\psline{->}(0,1)(0.55,2.1)
\uput[-90](0,1){$\scriptstyle A$}
\psline[linewidth=.4pt](2,4)(2.5,5)
\psline[linewidth=.4pt]{->}(1.5,3)(2.07,4.15)
\uput[90](2.5,5){$\scriptstyle B$}
\psline[linewidth=.4pt](2,2)(1.5,3)
\psline[linewidth=.4pt]{->}(2.5,1)(1.97,2.05)
\uput[-90](2.5,1){$\scriptstyle B$}
\pscircle[fillstyle=hlines*,fillcolor=white](1.25,3){0.5}
\endpspicture
&=& 
\pspicture[0.48](-1.5,0.7)(3.5,6)
\psline(0,3)(0,5)
\psline{->}(0,1)(0,3.15)
\psline[linewidth=.4pt](2,3)(2,5)
\psline[linewidth=.4pt]{->}(2,1)(2,3.1)
\endpspicture
+ 
\pspicture[0.5](-1,0.7)(3.5,7)
\psline(0.25,5.75)(0,6.5)
\psline{->}(0.5,5)(0.22,5.85)
\psline(0.68,3.95)(0.5,4.5)
\psline{->}(1,3)(0.65,4.05)
\psline(0.5,2)(1,3)
\psline{->}(0,1)(0.55,2.1)
\psline[linewidth=.4pt](2.25,5.75)(2.5,6.5)
\psline[linewidth=.4pt]{->}(2,5)(2.27,5.8)
\psline[linewidth=.4pt](1.8,3.9)(2,4.5)
\psline[linewidth=.4pt]{->}(1.5,3)(1.83,4.0)
\psline[linewidth=.4pt](2,2)(1.5,3)
\psline[linewidth=.4pt]{->}(2.5,1)(1.97,2.05)
\psframe[fillstyle=vlines](0.5,4.5)(2,5)
\pscircle[fillstyle=hlines*,fillcolor=white](1.25,3){0.5}
\endpspicture \\[-2mm]
&=&
\pspicture[0.48](-1.5,0.7)(3.5,6)
\psline(0,3)(0,5)
\psline{->}(0,1)(0,3.15)
\psline[linewidth=.4pt](2,3)(2,5)
\psline[linewidth=.4pt]{->}(2,1)(2,3.1)
\endpspicture
+ 
\pspicture[0.48](-1,0.7)(3.5,6.5)
\psline(0.25,4.5)(0,5.5)
\psline{->}(0.5,3.5)(0.23,4.6)
\psline[linewidth=.4pt](2.25,4.5)(2.5,5.5)
\psline[linewidth=.4pt]{->}(2,3.5)(2.27,4.55)
\psline(0.25,2)(0.5,3)
\psline{->}(0,1)(0.29,2.15)
\psline[linewidth=.4pt](2.25,2)(2,3)
\psline[linewidth=.4pt]{->}(2.5,1)(2.22,2.1)
\psframe[fillstyle=vlines](0.5,3)(2,3.5)
\endpspicture
+ 
\pspicture[0.48](-1,0.7)(3.5,7)
\psline(0.25,5.25)(0,6)
\psline{->}(0.5,4.5)(0.2,5.4)
\psline[linewidth=.4pt](2.25,5.25)(2.5,6)
\psline[linewidth=.4pt]{->}(2,4.5)(2.28,5.35)
\psline(0.5,3.5)(0.5,4)
\psline{->}(0.5,3)(0.5,3.65)
\psline[linewidth=.4pt](2,3.5)(2,4)
\psline[linewidth=.4pt]{->}(2,3)(2,3.6)
\psline(0.25,1.75)(0.5,2.5)
\psline{->}(0,1)(0.3,1.9)
\psline[linewidth=.4pt](2.25,1.75)(2,2.5)
\psline[linewidth=.4pt]{->}(2.5,1)(2.22,1.85)
\psframe[fillstyle=vlines](0.5,4)(2,4.5)
\psframe[fillstyle=vlines](0.5,2.5)(2,3)
\endpspicture + \hspace{5mm} \ldots
\et
\caption{An exact relation for the 4--point function, together with its
iterative solution in powers of the $AB$--irreducible kernel (represented
as a hatched box).}
\label{fig:dseq}
\end{figure}

\begin{figure}
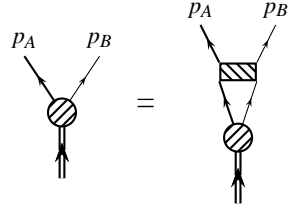

\pspicture[0.45](-0.75,3)(3.25,7.5)
\psline(0.75,5.75)(0.25,6.5)
\psline{->}(1.25,5)(0.59,6)
\uput[90](0.25,6.5){$\scriptstyle p_A$}
\psline[linewidth=.4pt](1.75,5.75)(2.25,6.5)
\psline[linewidth=.4pt]{->}(1.25,5)(1.88,5.95)
\uput[90](2.25,6.5){$\scriptstyle p_B$}
\psline[doubleline=true](1.25,4)(1.25,5)
\psline[doubleline=true,arrowscale=0.8]{->}(1.25,3.25)(1.25,4.2)
\pscircle[fillstyle=hlines*,fillcolor=white](1.25,5){0.4}
\endpspicture
= \pspicture[0.45](-0.75,3)(3.25,9)
\psline(0.5,7.5)(0.25,8)
\psline{->}(0.75,7)(0.42,7.65)
\uput[90](0.25,8){$\scriptstyle p_A$}
\psline[linewidth=.4pt](2,7.5)(2.25,8)
\psline[linewidth=.4pt]{->}(1.75,7)(2.05,7.6)
\uput[90](2.25,8){$\scriptstyle p_B$}
\psframe[fillstyle=vlines](0.75,6.5)(1.75,7)
\psline(0.95,5.9)(0.75,6.5)
\psline{->}(1.25,5)(0.87,6.1)
\psline[linewidth=.4pt](1.55,5.9)(1.75,6.5)
\psline[linewidth=.4pt]{->}(1.25,5)(1.6,6.05)
\psline[doubleline=true](1.25,4)(1.25,5)
\psline[doubleline=true,arrowscale=0.8]{->}(1.25,3.25)(1.25,4.2)
\pscircle[fillstyle=hlines*,fillcolor=white](1.25,5){0.4}
\endpspicture 
\caption{Diagrammatic representation of the Bethe--Salpeter equation.
The propagators with momenta $p_A$ and $p_B$ have to be 
understood as amputated, whereas they are contained in $\psi (p_A, p_B)$.}
\label{fig:bseq}
\end{figure}

The first equation that was devised to cope with this problem
is the Bethe--Salpeter equation \cite{BS51}. Its derivation starts
from a field theoretical identity for the 4--point function, the 
Dyson--Schwinger equation represented in Fig.\ \ref{fig:dseq}.
The kernel is given by the sum
of all $AB$--irreducible diagrams, i.e., diagrams that cannot be separated 
into two disconnected parts by cutting one internal $A$-- and one
$B$--propagator in any possible way. The identity becomes obvious after
solving it iteratively, which yields an expansion in powers of the kernel,
also represented in Fig.\ \ref{fig:dseq}.
A pole at $P^2 = M^2$ of the 4--point function can then be shown to exist
if and only if there exists a solution to the (homogeneous)
Bethe-Salpeter equation depicted in Fig.\ \ref{fig:bseq} (compare with 
Figs.\ \ref{fig:bspole} and \ref{fig:dseq}), with 
$P^2 = (p_A + p_B)^2 = M^2$.

In practice, calculating the kernel exactly is as impossible as
calculating the 4--point function itself, and the practical value of the
Bethe--Salpeter equation consists in suggesting an approximation scheme
appropriate for bound state calculations, in other words,
it suggests a choice of a subset of Feynman diagrams to be summed up
through an approximation of the kernel, and at the same time it devises
a method for its summation, namely the integration of the equation with
the approximate kernel. 

Due to the complexity in solving 4--dimensional integral
equations, virtually all equations that actually {\em have}\/ 
been solved use the so--called ladder approximation, where the kernel is 
replaced by the one--particle exchange diagram. The iterative solution of
the Dyson--Schwinger equation for the 4--point function
(Fig.\ \ref{fig:dseq}) in this approximation then leads to a sum of ladder
diagrams, with the propagators of the exchanged particles as rungs. In
addition, all self--energy corrections are neglected. In the
simplest case of three different scalar particles with cubic couplings,
the Bethe--Salpeter equation becomes explicitly
\be
\lefteqn{\left( p_A^2 - m_A^2 \right) \left( p_B^2 - m_B^2 \right)
\psi (p_A, p_B)} \hspace{5mm} \nonumber \\[1mm]
&=& i g^2 \int \frac{d^4 p_A'}{(2 \pi)^4} \frac{d^4 p_B'}{(2 \pi)^4} \,
\frac{(2 \pi)^4 \delta (p_A + p_B - p_A' - p_B')}
{(p_A - p_A')^2 - \mu^2 + i \epsilon} \, \psi (p_A', p_B') \:,
\ee
where $\mu$ is the mass of the exchanged particle and $g$ the coupling
constant.

Unfortunately, the ladder approximation has a number of well--known defects.
To begin with, the covariant 4--dimensional formulation implies a dynamics
in the relative energy $p^0 = p_A^0 - p_B^0$ which is related to the 
existence of so--called abnormal solutions \cite{Nak69,WC54}. 
These solutions have no non--relativistic
counterparts and appear to be inappropriate as solutions of the bound state
problem. In the one--body limit $m_B \to \infty$, the
Bethe--Salpeter equation in the ladder approximation does {\em not}\/
reduce to a relativistic equation for particle $A$ in the field generated
by the fixed source $B$, as is dictated by physical considerations 
\cite{WC54,Gro82}. Furthermore, gauge invariance and $s$--$u$ crossing 
symmetry are violated.

It is also known that, to remedy the latter three defects, besides the
one--particle exchange all crossed ladder diagrams have to be included 
in the kernel, 
defined as the set of all diagrams in which several bosons are exchanged
between the two constituents in an $AB$--irreducible way (the simplest
example being the crossed box diagram). Needless to say, this infinite
subset of diagrams cannot be summed exactly.

Together with the technical difficulties in resolving the 4--dimensional
equation, these problems have lead to the development of so--called
quasi--potential equations where the relative energy $p_0$ is integrated
over in such a way that covariance is conserved (although not manifestly). 
While this leads to 3--dimensional equations and eliminates the 
possibility of abnormal solutions, a consideration of
the pole structure of the ladder and crossed ladder diagrams at the same time 
allows to approximately sum up all these contributions to the
kernel. The crucial observation is that the crossed ladder diagrams
approximately {\em cancel}\/ some of the poles of the ladder diagrams, so
that an improved kernel can be obtained by taking into account {\em less}\/
residues in the $p_0$--integration \cite{Gro82,Gro69}. 

Out of the infinity of possible quasi--potential equations, I will 
cite the three most popular choices, all of which are consistent in
the one--body limit where some of the pole cancellations become exact. The
first quasi--potential equation to be formulated was the 
Blancenbecler--Sugar--Logunov--Tavkhelidze (BSLT) equation \cite{LTB63}.
It is the ``minimal'' choice in taking two--particle unitarity into
account via a dispersion relation. In the c.m.s. and for equal constituent
masses, the equation reads in the scalar model theory,
\be
\left[ 4 ( {\bf p}^2 + m^2 ) - E^2 \right] \psi ({\bf p})
- g^2 \int \frac{d^3 q}{(2 \pi)^3} \, \frac{1}
{\sqrt{ {\bf q}^2 + m^2}} \, \frac{\psi ({\bf q})}
{({\bf p} - {\bf q})^2 + \mu^2} = 0
\ee
with the lowest--order approximation to the kernel. I have put the equation in 
a form that resembles the non--relativistic Schr\"odinger equation, and 
$E = M = \sqrt{P^2}$.

The next quasi--potential equation to be presented is the Gross (or spectator)
equation \cite{Gro69}. It is obtained by putting one of the constituents on
its mass shell by hand, and reads for the scalar model considered before,
\be
\lefteqn{E \left( 2 \sqrt{{\bf p}^2 + m^2} - E \right)
\psi ({\bf p}) - g^2 \int \frac{d^3 q}{(2 \pi)^3} \, \frac{1}
{2 \sqrt{ {\bf q}^2 + m^2}}} \hspace{2cm} \nonumber \\[1mm]
& & {}\times \frac{\psi ({\bf q})}
{({\bf p} - {\bf q})^2 - \big( \sqrt{{\bf p}^2 + m^2}
- \sqrt{{\bf q}^2 + m^2} \,\big)^2 + \mu^2} = 0 \:. 
\ee
The difference of kinetic energies appearing in the denominator of the
potential term has its origin in the retardation of the interaction through
scalar boson exchange.

Finally, I cite the equal--time equation \cite{MW87}, which 
results from neglecting the relative energy in the kernel of the 
Bethe--Salpeter equation, and in addition takes the 
contribution of a pole from the crossed box diagram into account. 
Explicitly,
\be
\lefteqn{ \left[ 4 ( {\bf p}^2 + m^2 ) - E^2 \right]
\psi ({\bf p})} \nonumber \\[2mm]
& & {}- \left[ 2 - \frac{E^2}{4 ({\bf p}^2 + m^2)} \right] 
g^2 \int \frac{d^3 q}{(2 \pi)^3} \, \frac{1}
{\sqrt{ {\bf q}^2 + m^2}} \, \frac{\psi ({\bf q})}
{({\bf p} - {\bf q})^2 + \mu^2} = 0 \:.
\ee

As the last method within the Lagrangian approach to bound states I will
discuss a different technique, inspired by the worldline formulation of QFT
rather than by perturbation theory. In this method,
the field--theoretical Euclidean functional integral for the 4--point function
is rewritten as a path integral over the trajectories of the $A$-- and
$B$--particles in the so--called Feynman--Schwinger representation. The
field corresponding to the exchanged boson can then be analytically
integrated out \cite{ST93}.

In the quenched approximation without closed $A$-- and $B$--loops,
and after neglecting self--energy corrections to the
$A$-- and $B$--propagators and vertex corrections, the 4--point function
$G(y_A, y_B; x_A, x_B)$ is given by the sum of all
ladder {\em and}\/ crossed ladder graphs (not only $AB$--irreducible ones).
It can be calculated numerically by discretization of the path integrals and 
Monte Carlo methods. The mass $M_0$ of the ground state can then be extracted
from the dependence of the 4--point function on $T = (y_A^0 + y_B^0 - x_A^0
- x_B^0)/2$ in the limit $T \to \infty$, while the dependence on $y_A - y_B$
in the same limit determines the (relative) Bethe--Salpeter wave function.
The numerical results establish, for the first time, a
benchmark for bound state calculations, at least as long as $A$-- and
$B$--loops, self energies and vertex corrections are neglected.

\begin{figure}
\unitlength1cm
\begin{picture}(15,6)
\put(0,-0.2){\psfig{figure=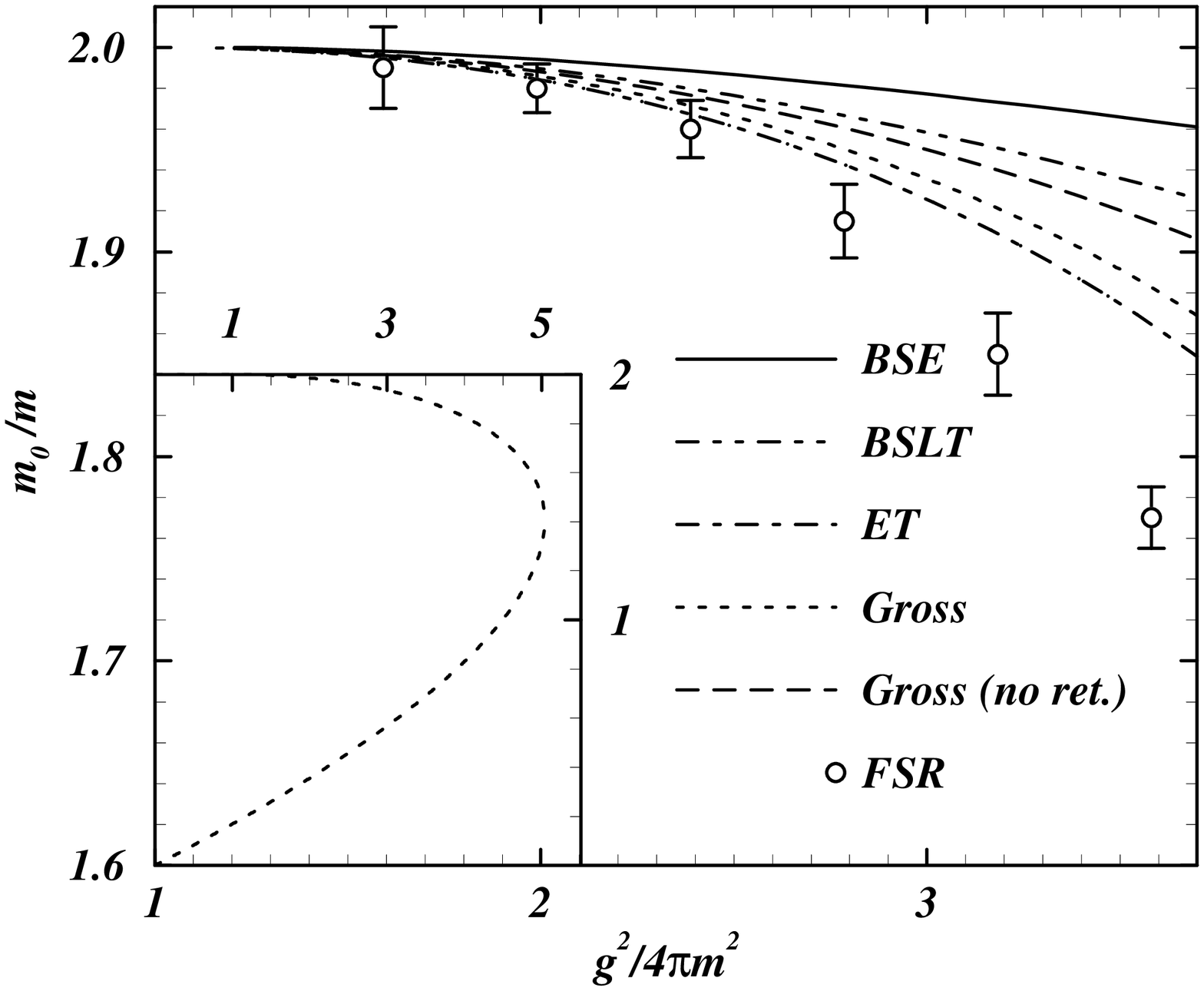,width=7cm}}
\put(7.5,-0.2){\psfig{figure=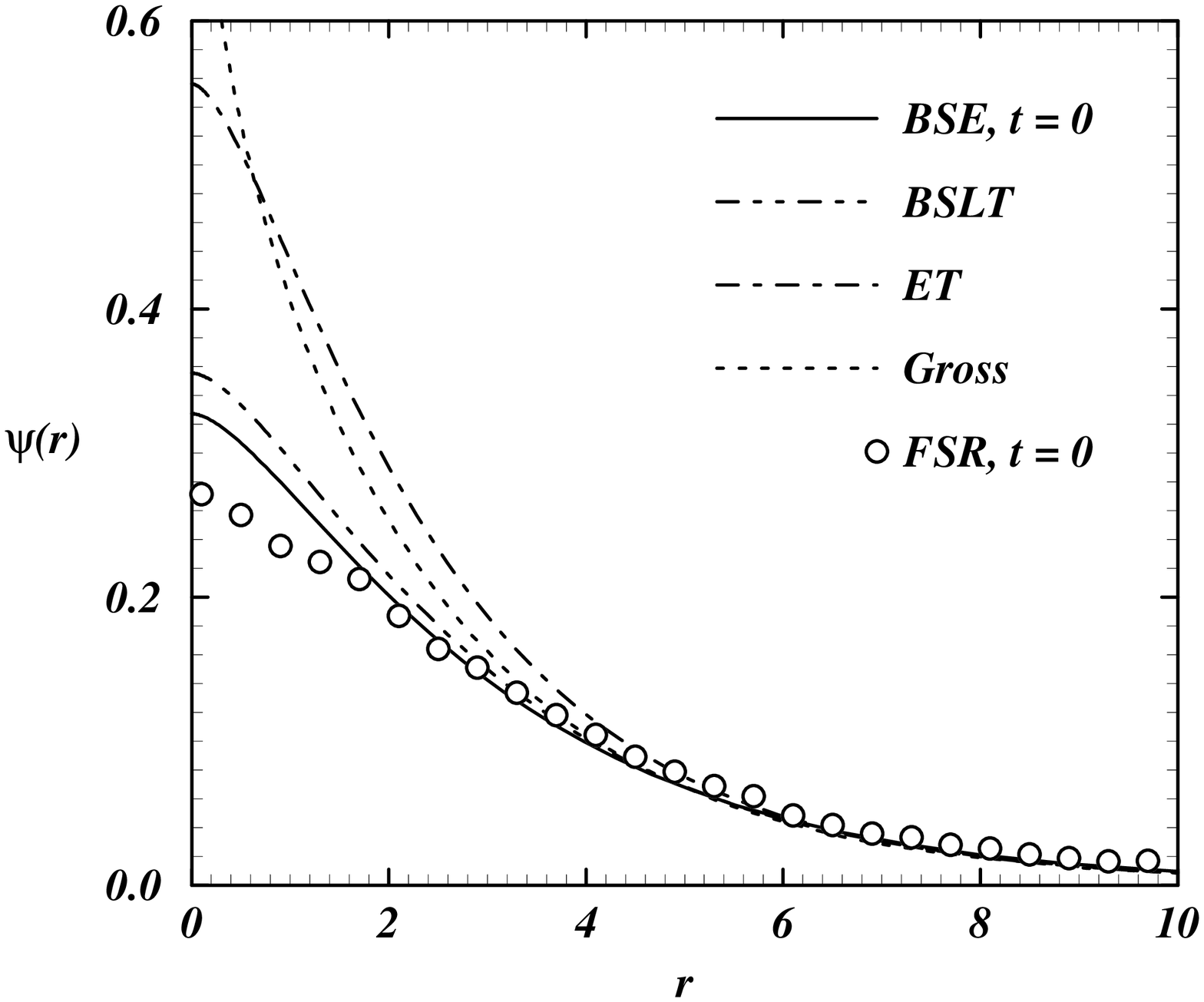,width=7cm}}
\end{picture}
\caption{The ground state mass $m_0$ as a function of the dimensionless 
coupling constant for $\mu/m = 0.15$, and the ground state (equal--time) 
wave function for
$m_0 = 1.882 m$; results from the Bethe--Salpeter equation (BSE), the
BSLT, equal--time (ET) and Gross (with and without retardation) 
quasi--potential equations, and from the Feynman--Schwinger representation 
(FSR) \cite{NT96}.}
\label{fig:gsres}
\end{figure}

The results for the ground state of the scalar model theory with mass ratio 
$\mu/m = 0.15$ are depicted in Fig.\ \ref{fig:gsres} for the Bethe--Salpeter 
equation, the BSLT, equal--time and Gross equations and from the 
Feynman--Schwinger representation \cite{NT96}.
Included is also a calculation with the Gross equation but {\em neglecting}\/
the retardation effects, and a representation of the unphysical branch of the
solution of the Gross equation (with retardation). It is seen that all
equations underestimate the binding energy (coming from ladder and crossed
ladder graphs), where the Bethe--Salpeter equation gives the ``worst'' and
the equal--time equation the relatively ``best'' results. The inclusion of 
retardation and the additional pole from the crossed box improve the results. 

Fig.\ \ref{fig:gsres} also compares the different results for the 
ground state wave functions for a fixed ground state mass of $m_0 = 1.882 m$
(and hence varying coupling constants). As far as the quality of 
the approximations is concerned, the outcome is precisely opposite to the
comparison of the binding energies. In conclusion, there is no compelling
evidence to prefer one of the quasi--potential equations over the others.

Turning now to the Hamiltonian approach to QFT, a bound state 
(or any scattering state, for that matter) is defined to be an eigenstate 
of the Hamiltonian $H$. The Hamiltonian is
decomposed into a free part $H_0$ and an interaction part $H_I$, where
the eigenstates of $H_0$ are well-known and used to construct the Fock
space ${\cal F}$ of the theory. It is assumed that any state in the 
interacting theory has a unique expansion in terms of the Fock space states.
Then the fundamental equation to be solved is
\be
H | \psi \rangle = E | \psi \rangle \:, \quad | \psi \rangle \in 
{\cal F} \:. \label{eq:schreq}
\ee
Now in general $H_I$ mixes the different $n$--particle sectors of ${\cal F}$,
so that Eq.\ \eqref{eq:schreq} is really an infinite system of coupled
(integral) equations for all sectors of Fock space, and the eigenstates
contain contributions from every sector (except where forbidden by
conservation laws). A systematic approximation scheme for the solution of 
Eq.\ \eqref{eq:schreq} is hence called for.

Most of the calculations in a Hamiltonian formalism are currently
done in QFT quantized on the light front (the 3--dimensional space
perpendicular to an arbitrarily chosen direction on the forward light
cone). The light--front dynamics differs from the usual equal-time quantized
theory in several aspects: the ``spectrum condition'' $p^+ > 0$ which is
related to the form of the light--front Hamiltonian, holds for real
{\em and}\/ virtual particles, and together with $p^+$--conservation at the 
vertices implies that there are no pair production processes in the vacuum.
However, the vacuum is, contrary to earlier folklore, in general 
{\em not trivial}\/ due to zero modes (constant field configurations 
on the light front). In particular, this seems to be the case in non--abelian
gauge theories \cite{BPP98}. Still, the structure of the vacuum should be
much simpler than in the usual formulation. Of equal importance is
the suppression of higher Fock components
by kinematical factors, even in the case of rather large coupling constants.
Obviously, these properties are particularly relevant to QCD. 

A serious problem is posed by the longitudinal divergencies which require 
non--local counterterms, i.e., entire
{\em functions}\/ of $p^+$ instead of a few numbers. A new, 
``non--perturbative'', renormalization theory has to be developed,
possibly along the lines of the coupling coherence scheme proposed by
Perry and Wilson \cite{PW93}.

The central idea behind all the approaches to the (approximate) solution of 
Eq.\ \eqref{eq:schreq} in light--front quantization is the ``reduction of
Fock space'' by similarity transformations, i.e., the block diagonalization
of the Hamiltonian converting \eqref{eq:schreq} into an 
(approximately) equivalent equation
on a small subspace of the Fock space ${\cal F}$. In practice, the
similarity transformations are usually realized as a sequence of unitary
transformations. Several similarity transformation schemes have been 
developed \cite{GWW94}. Among the few applications
of this technique to given field theories I mention the calculation of the
fine and hyperfine structure of positronium for a larger fine structure 
constant $\alpha = 0.3$ in the different schemes \cite{JPG97}. 
The numerical results are satisfactory, however,
a (quantitatively small) violation of rotational invariance and a
dependence on the similarity transformation scheme remain. The situation
is much less clear for bound state calculations in Yukawa theory because of
renormalization problems \cite{BPP98,GHP93}.

Given the difficulties with theories quantized on the light--front,
one might wonder whether
the similarity transformation method for the solution of Eq.\ \eqref{eq:schreq}
could not be applied successfully to the usual equal--time quantized theory.
In fact, a related formalism is currently being developed. It starts
from the following generalization of the Gell-Mann--Low theorem in QFT
\cite{GL51}: the time evolution determined by the Hamiltonian $H(t) = H_0 +
e^{- \epsilon |t|} H_I$ in the adiabatic limit $\epsilon \to 0$, maps
eigenstates of $H_0$ at $t \to -\infty$ that are isolated in the spectrum
to eigenstates of $H$ at $t = 0$. The same time evolution operator can be
shown to map $H_0$--invariant linear subspaces of ${\cal F}$ to $H$--invariant
subspaces under more general conditions \cite{Web00}. The Hamiltonian can
then be similarity transformed to the $H_0$--invariant subspace. For
example, if one considers the subspace of all free 2--particle states, 
then Eq.\ \eqref{eq:schreq} is equivalent to a quantum--mechanical 2--particle 
Schr\"odinger equation (on the subspace).

Applied to the scalar model considered before, the similarity transformation
leads to second order in $g$ to the following effective 
Schr\"odinger equation for $AB$--bound states \cite{WL02}:
\be
\lefteqn{\left( 2 \sqrt{{\bf p}^2 + m^2} - E \right)
\psi ({\bf p}) - g^2 \int \frac{d^3 q}{(2 \pi)^3} \,
\frac{1}{2 \sqrt{ {\bf p}^2 + m^2} \, 2 \sqrt{ {\bf q}^2 + m^2}}}
\nonumber \\[1mm]
& & {}\times \frac{1}{\sqrt{({\bf p} - {\bf q})^2 + \mu^2}}
\frac{\psi ({\bf q})}{\sqrt{{\bf p}^2 + m^2} - \sqrt{{\bf q}^2 + m^2} +
\sqrt{({\bf p} - {\bf q})^2 + \mu^2} } = 0 \:. \label{eq:ggleq}
\ee
In the hermitian version that was used by Wilson \cite{Wil70} and by 
Gl\"ockle and M\"uller \cite{GM81}, the equation becomes
\be
\lefteqn{\left( 2 \sqrt{{\bf p}^2 + m^2} - E \right)
\psi ({\bf p}) - g^2 \int \frac{d^3 q}{(2 \pi)^3} \,
\frac{1}{2 \sqrt{ {\bf p}^2 + m^2} \, 2 \sqrt{ {\bf q}^2 + m^2}}}
\nonumber \\[1mm]
& & {}\times \frac{\psi ({\bf q})}
{({\bf p} - {\bf q})^2 - \bigg( \sqrt{{\bf p}^2 + m^2}
- \sqrt{{\bf q}^2 + m^2} \bigg)^2 + \mu^2} = 0 \:.
\ee

\begin{figure}
\resizebox{.5\textwidth}{!}{\includegraphics{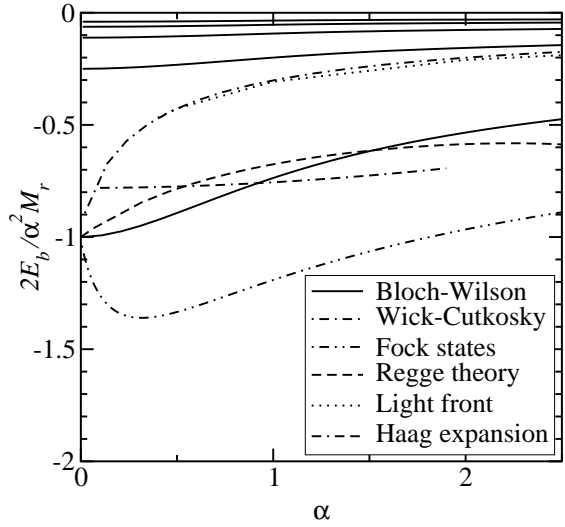}}
\caption{The binding energy $E_b = M - 2m$ of the ground state
and several excited $s$--states as a function of the ``fine structure
constant'' $\alpha = g^2/(16 \pi m^2)$ for $\mu = 0$, 
normalized to the non--relativistic ground state energy \cite{WL02}
($M_r = m/2$ is the reduced mass); results from the effective Schr\"odinger
equation \eqref{eq:ggleq} (``Bloch--Wilson''), the ladder
approximation of the Bethe--Salpeter equation (``Wick--Cutkosky''), a
straightforward Fock space expansion \cite{LB01}, the leading Regge
trajectory \cite{WLS01}, the Bethe--Salpeter equation in covariant 
light--front dynamics \cite{MC00}, and the Haag expansion \cite{GRS95}.}
\label{fig:ebres}
\end{figure}

A comparison of the numerical results for the binding energy of the 
ground state from the solution of Eq.\ \eqref{eq:ggleq} with those of the 
Bethe--Salpeter equation and several 
other equations which have not been discussed here for different reasons, is 
presented in Fig.\ \ref{fig:ebres} for the case $\mu = 0$ \cite{WL02}.
Solutions of Eq.\ \eqref{eq:ggleq} for 
some excited $s$--states are also shown. The ladder approximation of the 
Bethe--Salpeter equation is known in this case as the Wick--Cutkosky model, 
and is exactly solvable \cite{WC54}. The results from Eq.\ \eqref{eq:ggleq}
lie roughly half--way between the Bethe--Salpeter and the non--relativistic
results (value $(-1)$ in Fig.\ \ref{fig:ebres}), which is reasonable in view
of the results from the Feynman--Schwinger representation shown in Fig.\
\ref{fig:gsres}.

\begin{theacknowledgments}
Financial support by Conacyt grant 32729--E and by CIC--UMSNH is gratefully
acknowledged. I am also grateful to Norbert Ligterink for discussions and
reading of the manuscript. 
\end{theacknowledgments}

\end{document}